\documentclass[prb,twocolumn,amsmath,amssymb,showpacs]{revtex4-1}
\usepackage{graphicx}

\def\beq{\begin{eqnarray}}
\def\eeq{\end{eqnarray}}
\def\beqa{\begin{eqnarray}}
\def\eeqa{\end{eqnarray}}

\begin{document}

\title{Short-ranged and short-lived charge-density-wave order and pseudogap features in underdoped cuprates superconductors
}

\author{Andr\'es Greco and Mat\'{\i}as Bejas}

\affiliation{
Facultad de Ciencias Exactas, Ingenier\'{\i}a y Agrimensura and
Instituto de F\'{\i}sica Rosario
(UNR-CONICET).
Av. Pellegrini 250-2000 Rosario-Argentina.
}

\date{\today}

\begin{abstract}
The pseudogap phase of high-$T_c$ cuprates is controversially 
attributed to preformed pairs or to a phase which coexists and competes with superconductivity.
One of the challenges is to develop theoretical and experimental studies 
in order to distinguish between both proposals. 
Very recently, researchers at Stanford  have reported [M. Hashimoto {\it et al.}, Nat. Phys. {\bf 6}, 414 (2010); 
R.-H. He {\it et al.}, Science {\bf 331}, 1579 (2011)]
angle-resolved photoemission spectroscopy experiments on Pb-Bi$2201$ supporting 
the point of view that the pseudogap is distinct from superconductivity and associated to a spacial symmetry breaking 
without long-range order. In this paper we show that many features reported by these experiments 
can be described in the framework of the $t$-$J$ model considering self-energy effects in the proximity to a 
$d$ charge-density-wave instability. 
\end{abstract}

\pacs{74.72.-h, 74.25.Jb, 71.10.Fd, 79.60.-i}
\maketitle

Although the solid state physics community agrees
on the existence of the pseudogap (PG) phase in underdoped high-$T_c$ superconductors,
the origin of the PG and its relation with superconductivity (SC) 
is not clear from experimental and theoretical studies.
Interpretations run from descriptions where the PG
is intimately related to SC to others where the PG
is distinct from SC and both phases compete.\cite{hufner08}
Thus, experimental and theoretical studies are of central interest for solving this puzzle.

Angle-resolved photoemission spectroscopy (ARPES) is a valuable tool for studying the PG phase.\cite{damascelli03}
In the pseudogap phase ARPES experiments show, in the normal state below a characteristic temperature $T^*$,   
Fermi arcs\cite{norman98,shi08,terashima07,yoshida09} (FAs) 
(centered along the zone diagonal) instead of the expected Fermi surface (FS).
Despite the general consensus about the existence of FAs, their main characteristics are controversial. 
Some experiments\cite{norman98,shi08} report that near the antinode exists a single gap which is nearly 
independent of temperature. Moreover, in the superconducting state the gap follows the expected $d$-wave behavior 
along the FS. In contrast, other experiments\cite{terashima07,yoshida09} show 
that the gap follows the $d$-wave behavior near the node but deviates upward approaching the antinode.
In our opinion the difference in the spectra reported by different groups can not only be attributed to the different
systems under study. For instance, similar samples at similar doping (see Refs.[\onlinecite{shi08}] and [\onlinecite{terashima07}]) also show 
the mentioned discrepancy. Although the reason for 
these differences is not clear, results of Refs.[\onlinecite{norman98}] and [\onlinecite{shi08}] 
suggest that FAs are 
tied to the preformed pair scenario, while results from Refs.[\onlinecite{terashima07}] and [\onlinecite{yoshida09}] claim that 
arcs are associated to an order which is distinct from SC.

Recently,
ARPES results on Pb-Bi$2201$ were reported (Refs.[\onlinecite{hashimoto10}] and [\onlinecite{ruihua2011}]) 
showing that the PG  phase is distinct
from SC and associated to a spacial symmetry breaking without long-range order.
Since these experiments challenge scenarios about the origin of the PG and FAs, their
theoretical support is of high and broad interest.

In the framework of the $t$-$J$ model at mean-field level 
the carriers show a band dispersion 
$\epsilon_{k}= -2 t_{eff} [\cos(k_x)+\cos(k_y)]-4 t'_{eff} \cos(k_x) \cos(k_y) - \mu$,
with $t_{eff}=\left( t \frac{\delta}{2} + rJ \right)$ and $t'_{eff}=t'\frac{\delta}{2}$.
$t$, $t'$ and $J$ are the  
hopping between nearest-neighbor, next-nearest-neighbor and the nearest-neighbor 
Heisenberg coupling, respectively. $\delta$ is the doping away from half-filling, 
$\mu$ is the chemical potential,  and $rJ$ is the $J$-driven hopping term. 
This mean-field homogeneous Fermi liquid (HFL) phase presents an instability  
to a $d$ charge-density-wave ($d$-CDW) state\cite{chakravarty01} 
or a flux phase\cite{affleck88} (FP)
below a characteristic temperature $T_{FP}$ which decreases
with increasing doping (see Ref.[\onlinecite{greco09}] and references their in).

\begin{figure}
\centering
\includegraphics[width=7cm]{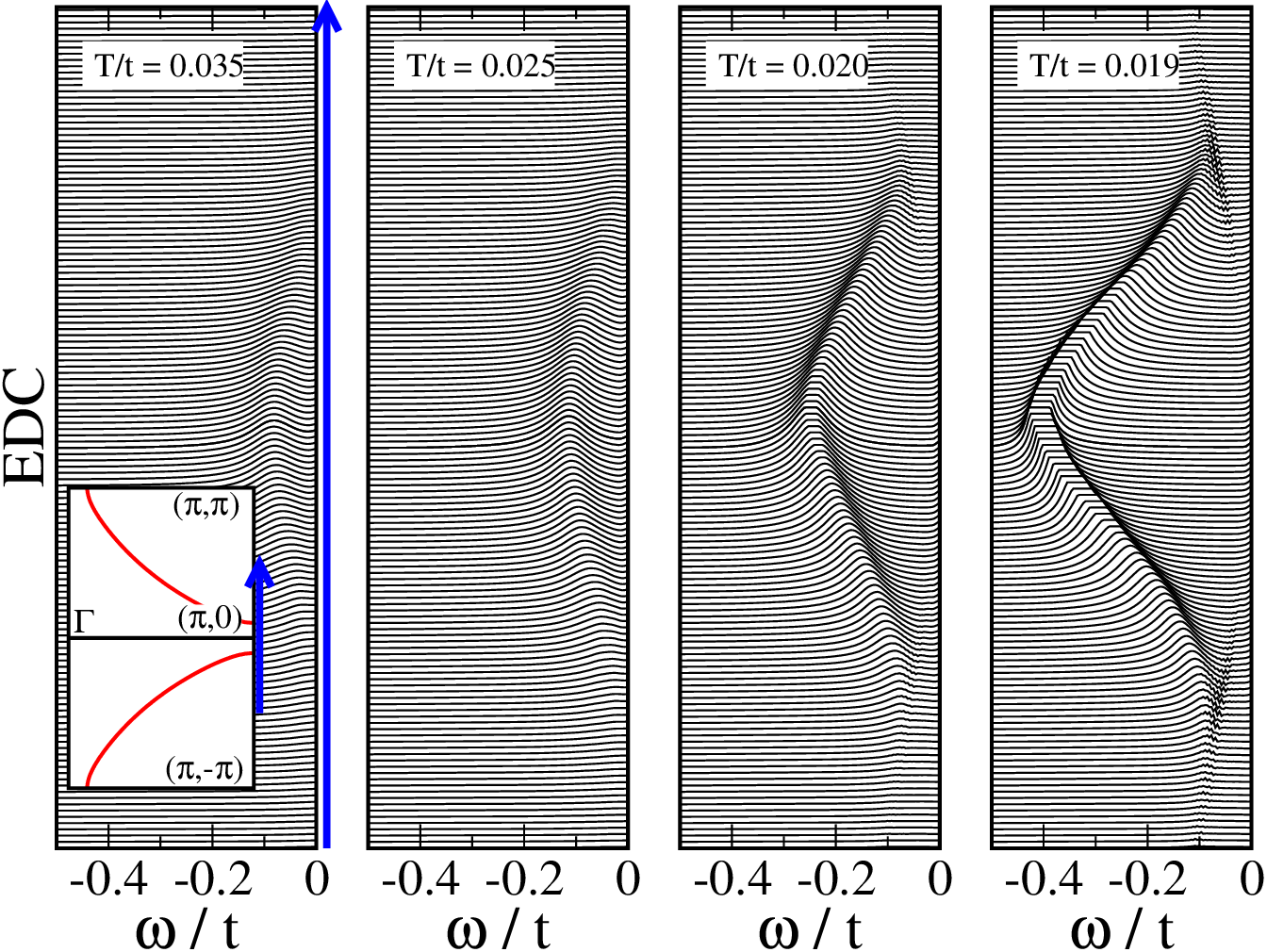}
\label{fig:imag}
\caption{(Color online)
Energy distribution curves in the normal state from a high temperature (left panel) to a lower temperature 
(right panel) near the $d$-CDW instability along the antinodal cut shown by an arrow in the inset of the left panel.  
}
\end{figure}
Beyond mean-field level, in the HFL phase and 
in the proximity to the $d$-CDW instability, the self-energy $\Sigma({\mathbf{k}},\omega)$ reads\cite{greco09}
\begin{eqnarray}\label{eq:SigmaIm0}
Im \, \Sigma({\mathbf{k}},\omega)&=&-\frac{1}{N_{s}}\sum_{{\mathbf{q}}} \gamma^2({\bf q},{\bf {k}}) Im \chi({\bf q},\omega-\epsilon_{{k-q}}) \nonumber\\
&& \hspace{-1cm}\times \left[n_{F}(-\epsilon_{{k-q}}) + n_{B}(\omega-\epsilon_{{k-q}})\right], \nonumber
\end{eqnarray}
\noindent 
where $n_B$  and $n_F$ are the Bose and Fermi factors, respectively, and $N_s$ is the number of sites.
$\chi({\bf q},\omega)=(\frac{\delta}{2})^2 [(8/J) r^2-\Pi({\bf q},\omega)]^{-1}$ is the $d$-CDW  susceptibility 
that diverges, at $\omega=0$ and ${\bf q}=(\pi,\pi)$,  and at $T_{FP}$. $\Pi({\bf q},\omega)$ 
is the electronic polarizability calculated with a form factor 
$\gamma({\bf q},{\bf k})=2 r [\sin(k_x-q_x/2)-\sin(k_y-q_y/2)]$. Note that since  ${\bf q} \sim (\pi,\pi)$ 
the form factor $\gamma({\bf q},{\bf k})$ transforms 
into $\sim [\cos(k_x)-\cos(k_y)]$ which indicates the $d$-wave character of the $d$-CDW.
Using the self-energy $\Sigma({\bf k},\omega)$,  the spectral 
function is calculated as usual.
This scenario leads to a PG and FAs which depend on doping and temperature in close agreement 
with the experiment, and occur via the coupling between quasiparticles (QPs) and the $d$-CDW soft mode 
in the proximity to the instability.\cite{greco09,greco08,bejas10} Here after $t'/t=-0.35$ and $J/t=0.3$. 
The hopping $t$ and the lattice constant $a$ of the square lattice are used as 
energy and length units, respectively.

\begin{figure}
\centering
\includegraphics[width=6.5cm]{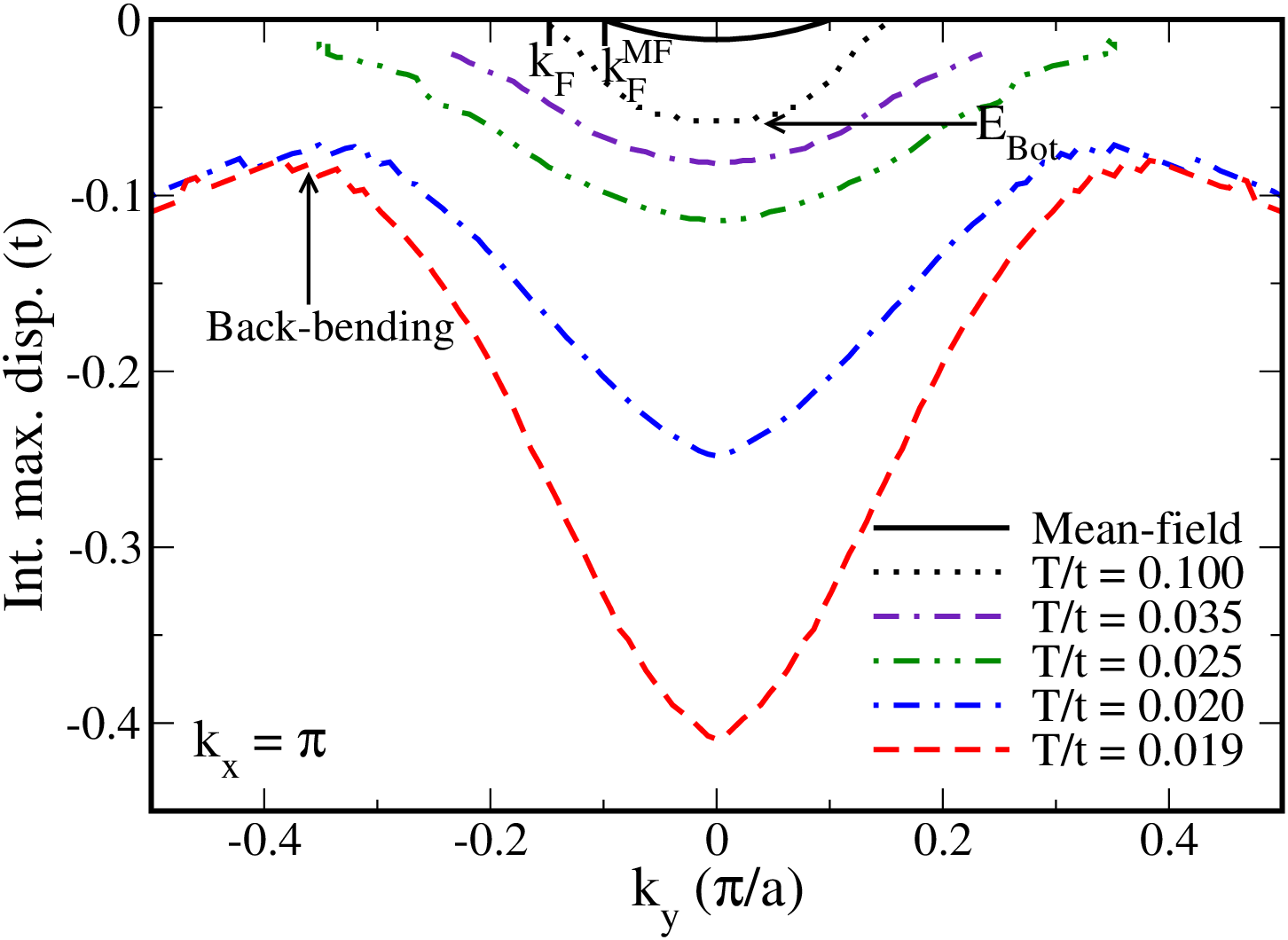}
\label{fig:imag}
\caption{(Color online)
Intensity maximum dispersion, for several temperatures in the normal state, taken along the antinodal cut. Solid line is
the predicted mean-field dispersion. At mean-field level the effective parameters are 
$t_{eff}/t\sim 0.08$ and  $t'_{eff}/t \sim -0.017$, while the effective parameters that reproduce the 
effective band at high temperature (dotted line) are $t_{eff}/t \sim 0.21$ and $t'_{eff}/t \sim -0.03$. 
Note that the Fermi wave vector ${\bf k_F}$ is different from the mean-field value ${\bf k_F^{MF}} \sim (\pi,0.1\pi)$. 
}
\end{figure} 
We show here that
many aspects observed in Refs.[\onlinecite{hashimoto10}] and [\onlinecite{ruihua2011}] can be discussed
in the framework of Refs.[\onlinecite{greco09}], [\onlinecite{greco08}], and [\onlinecite{bejas10}] showing 
that a theory where the PG phase is
distinct from  SC and associated with short-range fluctuations in the proximity
to a symmetry breaking instability is consistent with the experiment.

Similar to Fig. 1 g-l of Ref.[\onlinecite{hashimoto10}], in Fig. 1 
we present energy distribution curves (EDC) for momenta along
the antinodal cut indicated in the inset.
With no loss of generality, results are given in the HFL phase for doping $\delta=0.10$
and for several temperatures from 
$T/t=0.035$ to $T/t=0.019$ close but above $T_{FP}/t \sim 0.018$. 
At high temperature ($T/t=0.035$), EDC resemble a parabolic-like dispersion
as expected for usual metals. Lowering temperature, 
a PG like feature becomes evident near the antinode and, at the same time, a 
back-bending occurs in the dispersion. 
For summarizing the main results of Fig. 1, in Fig. 2, we show (similar to Fig. 1n in Ref.[\onlinecite{hashimoto10}]) the
intensity maximum dispersion for several temperatures. 
At high temperature ($T/t \sim 0.1$), no PG is observed  near the antinode and a parabola-like 
dispersion (dotted line), which crosses $\omega=0$ at ${\bf k_F} \sim (\pi,0.16 \pi)$, is obtained.
In the figure, we have also 
plotted results for the mean-field band (solid line). 
With decreasing temperature a PG opens near the antinode and, as in the experiment, a back-bending (marked by an arrow) 
is observed beyond ${\bf k_F}$,  in addition,  the energy of the 
band bottom ($E_{Bot}$) at ($\pi,0$) moves deeply to lower binding energies.
In Ref.[\onlinecite{hashimoto10}], the existence of a back-bending beyond ${\bf k_F}$ was discussed as a result 
that rules out preformed pairs and favors 
a scenario where the PG is distinct from SC. Present results are consistent with that interpretation as discussed below.

\begin{figure}
\centering
\includegraphics[width=6.5cm]{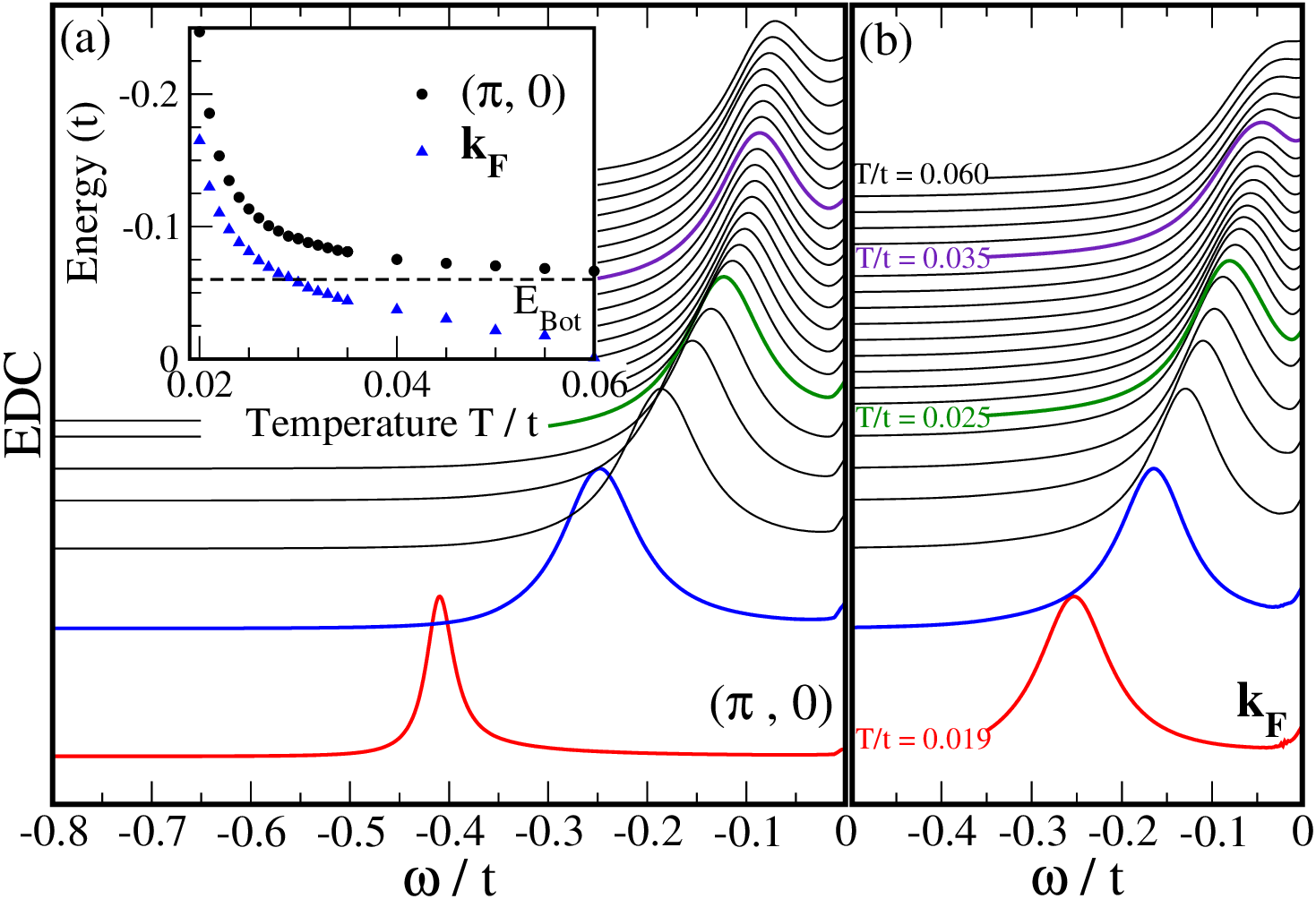}
\label{fig:imag}
\caption{(Color online)
(a) Energy distribution curves for momentum ($\pi,0$) for several temperatures. (b) The same as (a) 
for momentum ${\bf k_F}$. The inset shows the energy position of the maximum of the energy distribution 
curves as a function of temperature for ${\bf k}=(\pi,0)$ (circles) and ${\bf k_F}$ (triangles).
}
\end{figure}
Following the comparison with results in Ref.[\onlinecite{hashimoto10}], in Figs. 3(a) and 3(b) we have plotted EDC at ($\pi,0$) 
and ${\bf k_F}$, respectively, for several temperatures.  
For ${\bf k}=(\pi,0)$, the energy of the maximum of the EDC 
moves, with increasing temperature, toward $E_{Bot}$ 
as shown by circles in the inset. 
Figure 3(b) shows that, with increasing temperature, the PG at ${\bf k_F}$ closes and, at the same time, fills as in the experiment. 
In the inset the energy of the maximum of the EDC at ${\bf k_F}$ is plotted (triangles) as a function of temperature showing that, 
for $T/t \gtrsim 0.06$ no PG is observed. Since  EDC at ${\bf k_F}$
show a maximum at $\omega=0$ for $T/t \gtrsim 0.06$, we identify 
the pseudogap temperature with $T^*/t \sim 0.06$. In addition, we also identify the
dotted line in Fig. 2  with the effective metallic band (recovered at high temperature)  
discussed in Ref.[\onlinecite{hashimoto10}]. 
Finally, we note that in the present theory the opening of the PG is not associated to an abrupt transition 
but to a smooth crossover.
\begin{figure}
\centering
\includegraphics[width=6.5cm]{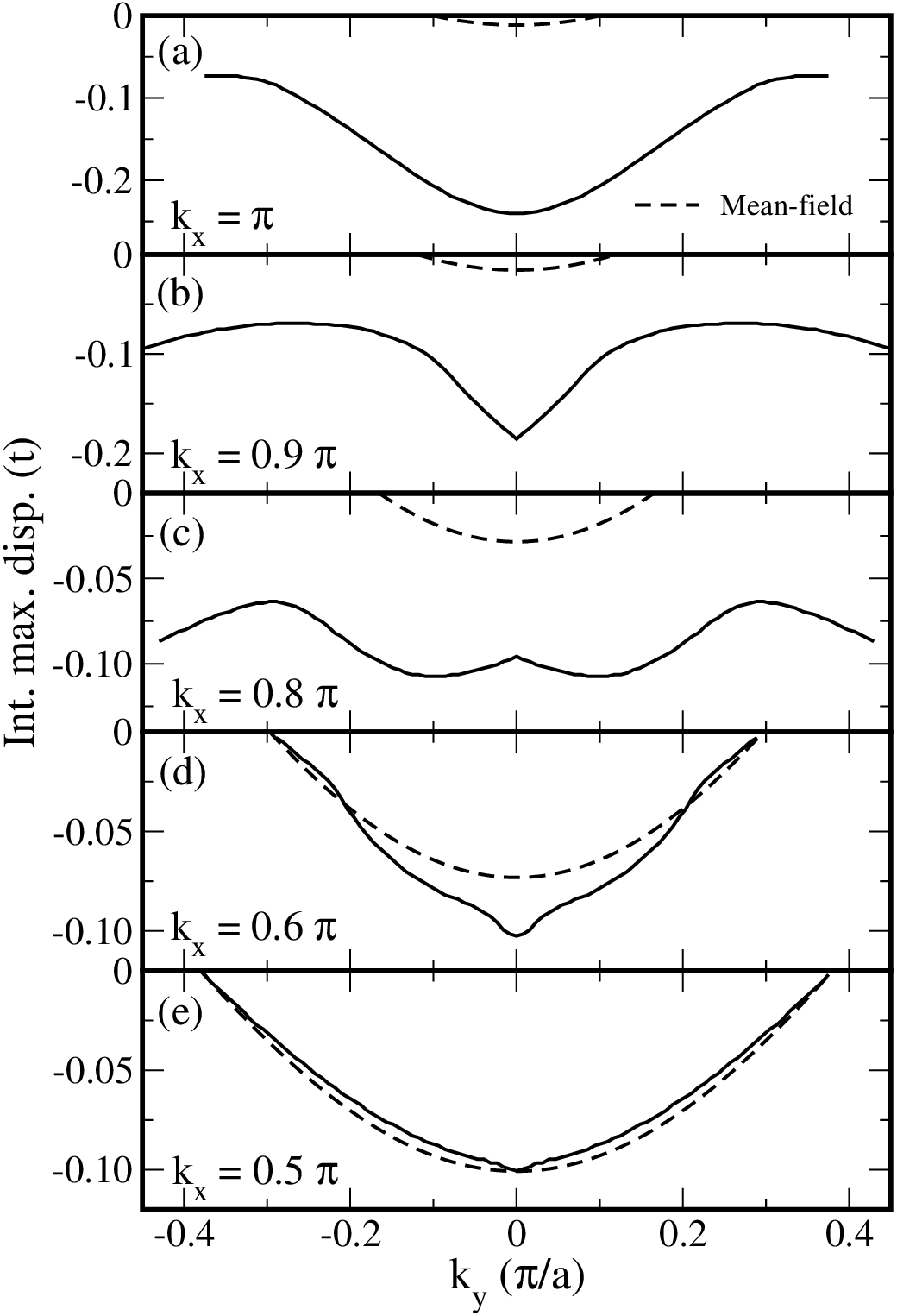}
\label{fig:imag}
\caption{
Intensity maximum dispersion for several cuts, parallel to the antinodal cut, at $T/t=0.020$. 
Dashed line in each panel is 
the predicted mean-field dispersion for the corresponding cut.  
}
\end{figure}

Using the accepted value $t=0.4$ eV, $T^* \sim 270$ K, $E_{Bot} \sim 25$ meV, and the energy at the 
back-bending $\sim 40$ meV. Since our description does not require phenomenological 
parameters as a value for the PG energy and  its temperature dependence, as in other approaches,\cite{leblanc10} 
the agreement with the experiment can be considered satisfactory.

Next, we show predictions of our theory for other cuts on the Brillouin zone.  
Similar to Fig. 2, in Fig. 4 we have plotted, for $T/t=0.020$, the intensity maximum dispersions for cuts parallel 
to the antinodal cut. In each cut, we fix $k_x$ to be $\pi, 0.9 \pi, 0.8 \pi, 0.6\pi $, and $0.5\pi $, and vary $k_y$.
While for cuts close to the antinode [($\pi,k_y$), ($0.9 \pi,k_y$), and ($0.8 \pi,k_y$)]
a PG and a back-bending are clearly observed, for cuts near the node  [($0.6\pi,k_y$) and ($0.5\pi,k_y$)]
the back-bending and the PG wash out and a FS crossing occurs. 
This behavior is consistent with the prediction of FAs by 
present theory.\cite{greco08,greco09,bejas10} 
In addition, different to the cuts near the antinode (see Figs. 1 and 3),  for cuts near the node 
the QP peaks (not shown) become sharper and the renormalized dispersion is 
closer to the bare mean-field band (dashed line). 
This fact reflects  that the effect of the $d$-CDW mode diminishes approaching the node.\cite{greco09}

\begin{figure}
\centering
\includegraphics[width=6.5cm]{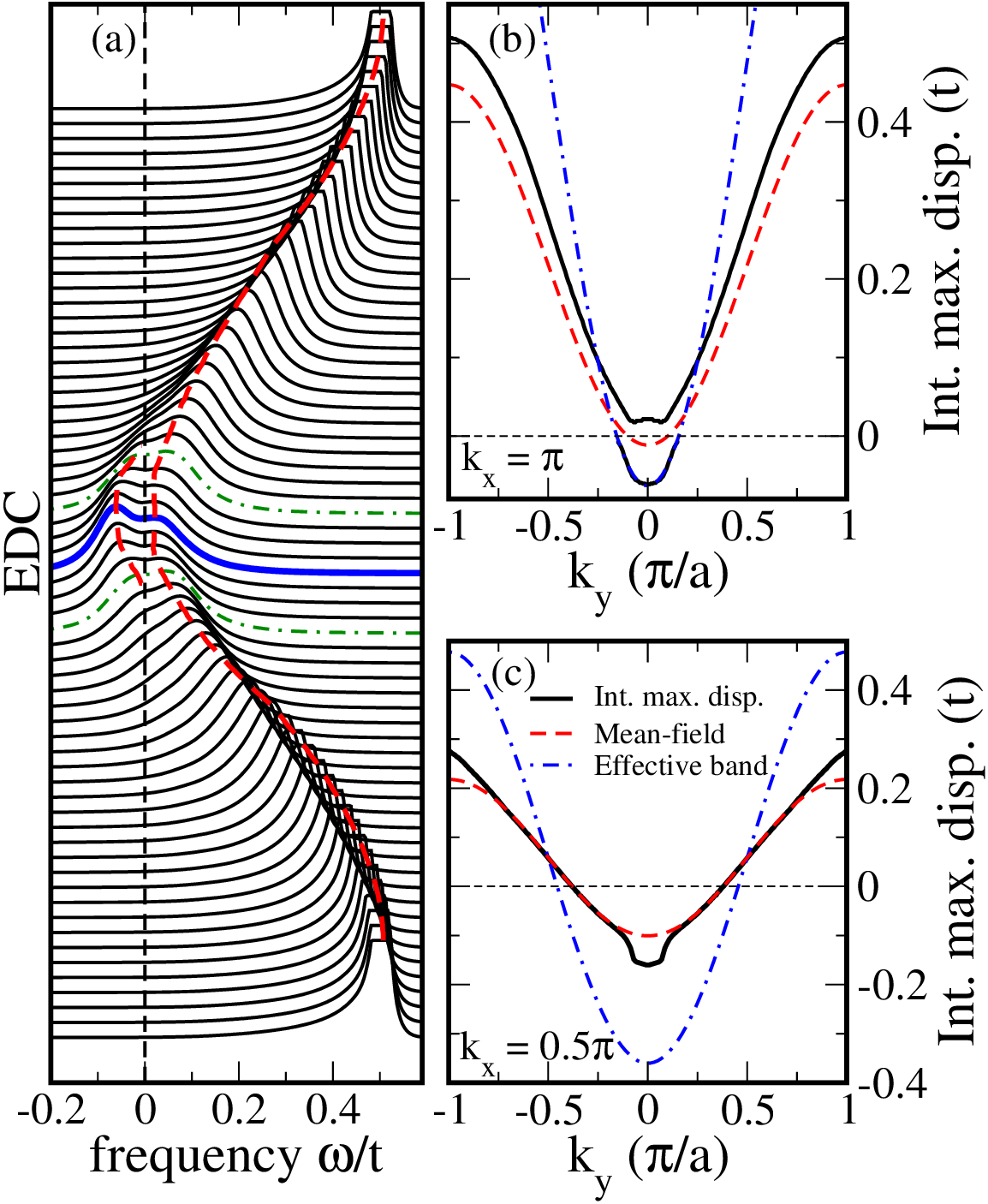}
\label{fig:imag}
\caption{(Color online)
(a) Energy distribution curves along the antinodal cut for $T/t=0.1$. 
Thick solid blue line is the energy distribution curve for ${\bf k}=(\pi,0)$. Dashed red lines represent 
a guide for the eye 
showing that the double peak structure at both sides of $\omega=0$ near ${\bf k}=(\pi,0)$ merges 
into only one peak away from ${\bf k}=(\pi,0)$, which follows the predicted mean-field result. 
Note that at ${\bf k_F}$ (dashed-dotted green line), no pseudogap is observed for this temperature. 
(b) Intensity maximum dispersion (solid line), the effective band (dashed-dotted line) discussed in text, and the mean-field 
dispersion (dashed line) along the antinodal cut for $T/t=0.1$ (c) The same as (b) for the cut ($0.5 \pi,k_y$).  
}
\end{figure}
It is worth to mention that a  generic long-range CDW phase may show, qualitatively, 
some of the features presented here such as, for instance, the back-bending beyond ${\bf k_F}$.
Some differences with the long-range CDW point of view are:
(a) While in a CDW phase the translational symmetry is broken, in the present theory, the system is 
always in the HFL phase but,  under the influence of short-range fluctuations.  
(b) While sharp QP peaks are expected in a CDW phase, our results show broad structures as in the experiment.
These broad structures are  intrinsic and not related to  the instrumental  resolution as suggested 
in Ref.[\onlinecite{leblanc10}].
For explaining the broad spectra observed in the experiment, in Ref.[\onlinecite{hashimoto10}] a static density 
wave order with a finite correlation length was assumed. In addition, it was suggested that the observed fast sink,
with decreasing temperature, of the band bottom implies an incommensurate character of the charge order. Both, 
the broad spectra and the temperature behavior of the band bottom are naturally 
described by the present picture. We think that this fact suggests, besides of a short correlation length, a slowly dynamical  
character of the charge order, and these conditions are fulfilled in the fluctuating regime near the instability. 
Since the discussion about the interpretation of the PG as a phase with long-range order versus a phase 
without long-range order is highly controversial (see Ref.[\onlinecite{vishik}] for a review),
we expect that our finding may contribute to the distinction between 
both scenarios. 
\begin{figure}
\centering
\includegraphics[width=6.5cm]{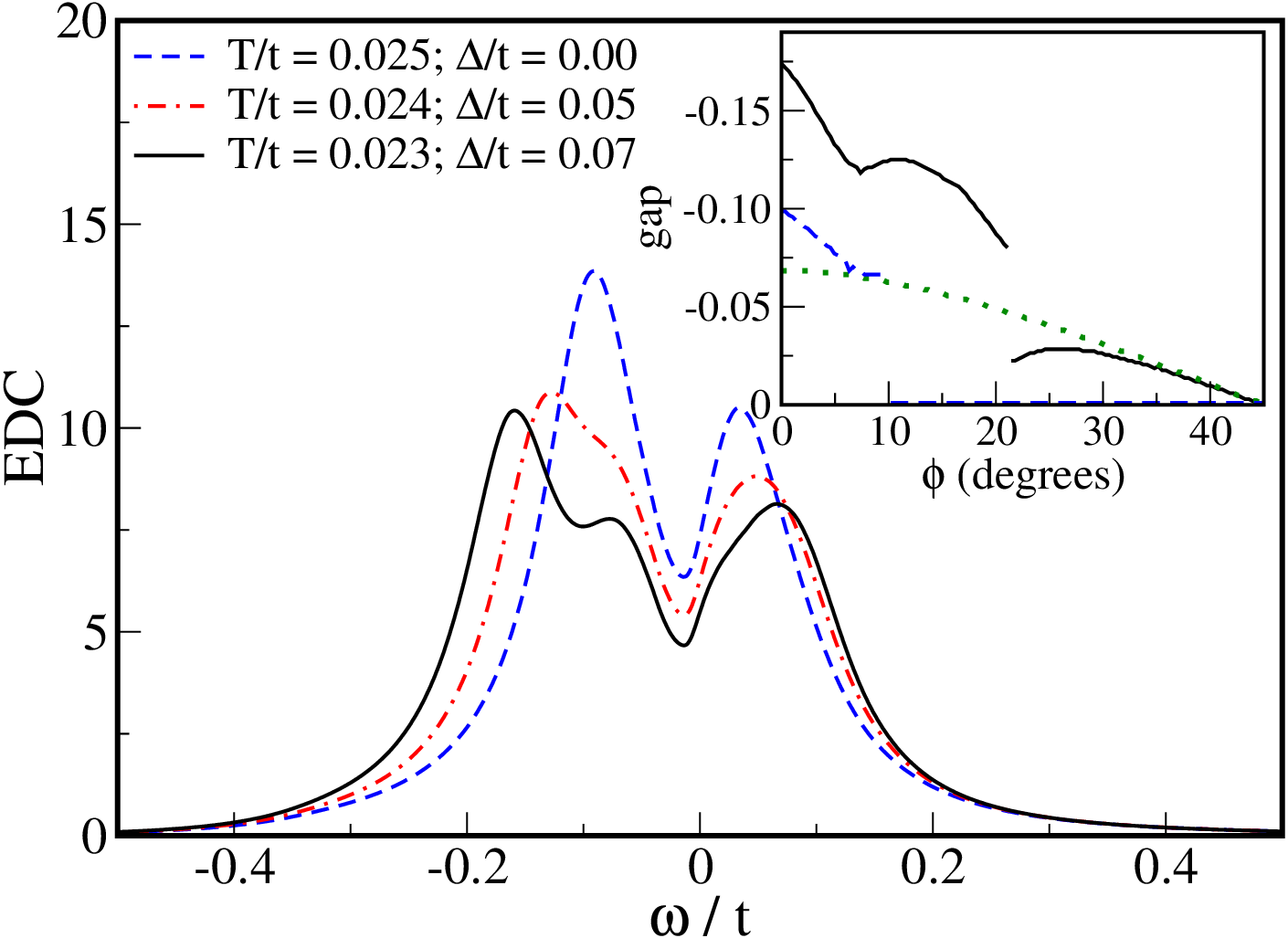}
\label{fig:imag}
\caption{(Color online)
EDC obtained for various superconducting gaps values $\Delta$ (see text). 
Inset: Energy position of the leading gap  of the EDC as a function of the FS angle $\phi$ for the normal state 
(dashed line) and in the superconducting state (solid line).}
\end{figure}

Now, we discuss some characteristics of the high temperature phase. In Figs. 5(a) and 5(b), we present EDC
and the corresponding intensity maximum dispersions, respectively, for the antinodal cut at $T/t=0.1$. 
Note the broad feature with structure at both sides of $\omega=0$ at $(\pi,0)$ [thick solid blue line in Fig. 5(a)].
This double peak structure  merges into only
one feature away from $(\pi,0)$ approaching the predicted mean-field band [dashed line in Fig. 5(b)]. 
The parabola-like band, discussed in Fig. 2
at high temperature, follows the effective band 
[dashed-dotted line in Fig. 5(b)] for $\omega <0$. 
We note that the closing of the PG at  $T/t \sim 0.06$
is mainly due to a filling effect which occurs from the merging of the broad peak at $\omega <0$ with 
the one at $\omega >0$. 
In Fig. 5(c), we have plotted, for the cut $(0.5\pi,k_y)$, the intensity maximum dispersion of the EDC at $T/t=0.1$.
Different to the antinodal cut, the intensity maximum dispersion is closer to the  
mean-field band (dashed line) for both, $\omega < 0$ and $\omega > 0$. 
From the experimental point of view, this result means that above $T^*$, a unique set of 
parameters might not describe 
the dispersion for all cuts, as it should be if the high temperature phase can be associated 
with a usual metal. It is interesting to see whether this  prediction may be checked by 
the experiment.

In  Ref.[\onlinecite{hashimoto10}], 
it was reported that with decreasing temperature, the spectra  and the leading edge gap near the antinode 
smoothly become broad and move 
toward higher binding energy, respectively,    
with no significant change below the superconducting critical temperature $T_c$. 
At first sight, it would expect that $d$-CDW
fluctuations would be washed out once superconductivity sets in.  
However, considering that the superconducting gap is smaller than the pseudogap, we can approximate 
the Green function in the superconducting state as
$G^{-1}_{SC}(k,i\omega) = G^{-1}(k,i\omega) + \Delta^2_k G(k,-i\omega)$, where $G(k,i\omega)$ is the Green 
function describing the spectral functions of  Fig. 3(b), and $\Delta_k = \Delta (\cos k_x - \cos k_y ) / 2$ 
is a $d$-wave superconducting gap. In Fig. 6, we present EDC at  ${\bf k_F}$.  
In the normal state ($\Delta=0$) (assumed to be the previous result for $T/t=0.025$), 
the peaks at both sides of $\omega=0$ are broad and asymmetric. With increasing $\Delta$ 
(decreasing temperature below $T_c$), the broad peak slowly moves to lower energies as in the experiment.\cite{hashimoto10} 
It is interesting to note the presence of an inner shoulder which is a direct indication of 
two distinct energy scales below $T_c$. A similar shoulder was discussed in Ref.[\onlinecite{hashimoto10}], and more 
clearly reported in Ref.[\onlinecite{ruihua2011}] after improving the experimental resolution. We think that an  unambiguous 
experimental confirmation of an inner shoulder (or a peak) 
below $T_c$ is of fundamental interest for solving the puzzle about one versus two energy scales. 
The reason for the absence of this shoulder in some ARPES experiments\cite{norman98,shi08} is an open question and out of the scope 
of the present paper.
In the inset of Fig. 6 we plot the energy position of the leading gap  of the EDC as a function of the FS angle $\phi$ from the antinode 
($\phi=0$) to the node ($\phi=45$ degrees). For $\Delta=0$ (dashed line), the region near the antinode is gaped leading to
Fermi arcs. For $\Delta/t=0.07$ (solid line), the curve follows the $d$-wave behavior (dotted line) only near the node, and deviates 
upward approaching the antinode. It is interesting to see that for Pb-Bi$2201$, the gap follows this trend in similar way to  
Refs.[\onlinecite{terashima07}] and [\onlinecite{vishik}] (see Fig. 4S in Ref.[\onlinecite{ruihua2011}]).

Concluding, we have shown that the scattering between carriers and $d$-CDW fluctuations, which occur in the 
proximity to the $d$-CDW instability, describes many features observed in recent ARPES experiments. 
The proposed scenario supports the point of view that the pseudogap is distinct from  superconductivity and 
associated with short-range and short-lived CDW fluctuations. 

The authors thank H. Parent, R.-H. He, and R. Zeyher for valuable discussions.


\begin{thebibliography}{10}


\bibitem{hufner08}
S. H\"ufner, M.A. Hossain, A. Damascelli, and G.A. Sawastzky, Rep. Prog. Phys. {\bf 71}, 062501 (2008). 
M. R. Norman, D. Pines and C. Kallin, Adv. Phys. {\bf 54}, 715 (2005).

\bibitem{damascelli03} A. Damascelli, Z. Hussain, and Z.-X.Shen,  Rev. Mod. Phys. {\bf 75}, 473 (2003).

\bibitem{norman98}
M.R. Norman {\it et al.}, Nature {\bf 392}, 157 (1998).
A. Kanigel {\it et al.}, Nature Physics {\bf 2}, 447 (2006).
A. Kanigel {\it et al.}, Phys. Rev. Lett. {\bf 99}, 157001 (2007).

\bibitem{shi08}
M. Shi  {\it et al.}, Phys. Rev. Lett. {\bf 101}, 047002 (2008).

\bibitem{terashima07}
K. Terashima {\it et al.}, Phys. Rev. Lett. {\bf 99}, 017003 (2007).

\bibitem{yoshida09} T. Yoshida {\it et al.}, Phys. Rev. Lett. {\bf 103}, 037004 (2009).
T. Kondo {\it et al.}, Phys. Rev. Lett. {\bf 98}, 267004 (2007).
T. Kondo {\it et al.}, Nature {\bf 457}, 296 (2009). 
K. Tanaka, {\it et al.}, Science {\bf 314}, 1910 (2006).  
W.S. Lee, {\it et al.}, Nature {\bf 450}, 81 (2007).  
J.-H. Ma  {\it et al.}, Phys. Rev. Lett. {\bf 101}, 207002 (2008).

\bibitem{hashimoto10}
M. Hashimoto {\it et al.}, Nature  Phys. {\bf 6}, 414 (2010).

\bibitem{ruihua2011}
R.-H. He, {\it et al.}, Science {\bf 331}, 1579 (2011).

\bibitem{chakravarty01}
S. Chakravarty, R. B. Laughlin, D. K. Morr, and C. Nayak, Phys. Rev. B {\bf 63}, 094503 (2001).

\bibitem{affleck88}
I. Affleck and J.B. Marston, Phys. Rev. B {\bf 37}, 3774 (1988).
E. Cappelluti and R. Zeyher, Phys. Rev. B {\bf 59}, 6475 (1999).

\bibitem{greco09}
A. Greco, Phys. Rev. Lett. {\bf 103}, 217001 (2009).

\bibitem{greco08}
A. Greco, Phys. Rev. B {\bf 77}, 092503 (2008).

\bibitem{bejas10}
M. Bejas, G. Buzon, A. Greco, and  A. Foussats, Phys. Rev. B {\bf 83}, 014514 (2011).

\bibitem{leblanc10}
J. P. F. LeBlanc, J. P. Carbotte, and E. J. Nicol, Phys. Rev. B {\bf 83}, 184506 (2011).

\bibitem{vishik}
I. M. Vishik {\it et al.}, New J. Phys. {\bf 12}, 105008 (2010).




\end{thebibliography}
\end{document}